\documentclass[a4paper,11pt]{article}
\pdfoutput=1 

\usepackage{jcappub} 

\title{\boldmath Probing cosmic anisotropy with GW/FRB as upgraded standard sirens}

\author[a,b]{Rong-Gen Cai,}
\author[a,b,1]{Tong-Bo Liu,\note{Corresponding author.}}
\author[c]{Shao-Jiang Wang,}
\author[a,b]{Wu-Tao Xu}

\affiliation[a]{CAS Key Laboratory of Theoretical Physics, Institute of Theoretical Physics, Chinese Academy of Sciences, P.O. Box 2735, Beijing 100190, China}
\affiliation[b]{School of Physical Sciences, University of Chinese Academy of Sciences, No.19A Yuquan Road, Beijing 100049, China}
\affiliation[c]{Tufts Institute of Cosmology, Department of Physics and Astronomy, Tufts University, 574 Boston Avenue, Medford, Massachusetts 02155, USA}

\emailAdd{cairg@itp.ac.cn}
\emailAdd{liutongbo@itp.ac.cn}
\emailAdd{schwang@cosmos.phy.tufts.edu}
\emailAdd{xwutao@itp.ac.cn}

\abstract{Recently it was shown that cosmic anisotropy can be well tested using either standard siren measurement of luminosity distance $d_\mathrm{L}(z)$ from gravitational-wave (GW) observation or dispersion measure ($\mathrm{DM}(z)$) from fast radio burst (FRB). It was also observed that the combined measurement of $d_\mathrm{L}(z)\cdot\mathrm{DM}(z)$ from the GW/FRB association system as suggested in some of FRB models is more effective to constrain cosmological parameters than $d_\mathrm{L}(z)$ or $\mathrm{DM}(z)$ separately. In this paper, we  show that this upgraded siren from the  combined GW/FRB observations could test cosmic anisotropy with a double relative sensitivity compared to the usual standard siren from GW observations only. }

\begin{document}
\maketitle
\flushbottom

\section{Introduction}
\label{sec:intro}

The power of the standard siren \cite{Schutz:1986gp,Holz:2005df} starts to be appreciated since the first discovery of binary neutron star (BNS) merger \cite{TheLIGOScientific:2017qsa,Abbott:2017xzu}. The luminosity distance can be directly determined from the gravitational waveform of coalescing binaries and the redshift can be inferred from the associated electromagnetic (EM) counterpart. In a previous work \cite{Cai:2017aea}, we have demonstrated the promiseful  perspective of constraint ability on cosmic anisotropy using gravitational wave (GW) as standard siren. We found that with a few hundreds standard siren events, the cosmic isotropy can be ruled out at $3\sigma$ confidence level (C.L.) and the dipole direction can be constrained within $10\sim20\%$ at 3$\sigma$ C.L. if the dipole amplitude is of order $\mathcal{O}(10^{-2})$. See \cite{Cai:2016sby,Cai:2017yww,Wang:2018lun,Zhang:2018byx,Wei:2018cov} for more recent studies on the standard siren and its application in cosmology and \cite{Cai:2017cbj} for a review and references therein.

On the other hand, it was indicated in some models \cite{Petroff:2016tcr,Platts:2018hiy} of fast radio burst (FRB) \cite{Lorimer:2007qn} that, the observed millisecond-duration radio transients could originate from the mergers of binary neutron stars (BNS) \cite{Totani:2013lia, Wang:2016dgs, Yamasaki:2017hdr}, black hole-neutron star (BHNS) \cite{Mingarelli:2015bpo},  double charged black holes (BHs) \cite{Zhang:2016rli} or charged and rotating BHs \cite{Liu:2016olx} (See also \cite{Falcke:2013xpa, Nathanail:2017wly, Most:2018abt} for the blitzar model of FRB from a single collapsing magnetised neutron star). A typical FRB could exhibit high dispersion measure (DM) of order $\mathcal{O}(10^2\sim10^3)\,\mathrm{pc}\cdot\mathrm{cm}^{-3}$ with negligible uncertainties of order $\mathcal{O}(10^{-2}\sim10^{-1})\,\mathrm{pc}\cdot\mathrm{cm}^{-3}$ \cite{Petroff:2016tcr}. Recently, the cosmic anisotropy was tested in \cite{Qiang:2019zrs} using dispersion measure for some simulated FRB events with known redshifts. With a few hundreds FRB events, the cosmic anisotropy can be found if the dipole amplitude is of order $\mathcal{O}(10^{-2})$. Other applications of FRBs on cosmology can be found in \cite{Deng:2013aga,Yang:2016zbm,Gao:2014iva,Zhou:2014yta,Yu:2017beg,Yang:2017bls,Wei:2018cgd,Li:2017mek,Jaroszynski:2018vgh,Madhavacheril:2019buy,Wang:2018ydd,Walters:2017afr}. 

In particular, an ungraded standard siren was proposed in \cite{Wei:2018cgd} when the binary system exhibits both GW and FRB signals. Combining the luminosity distance $d_{\rm L}(z)$ from GW and dispersion measures $\mathrm{DM}_\mathrm{IGM}(z)$ from FRB through intergalactic medium (IGM), the authors found that it could constrain the cosmological parameters more strongly than that using separately either $d_{\rm L}(z)$ or $\mathrm{DM}_\mathrm{IGM}(z)$.  It was also discussed that the FRB event rate is around $1.4\times 10^{3} \rm Gpc^{-3}yr^{-1}$ \cite{Zhang:2016rli}, meanwhile the BNS merger rate is estimated to be $1540_{-1220}^{+3200}\rm Gpc^{-3}yr^{-1}$\cite{TheLIGOScientific:2017qsa}, which are at the same order of magnitude. It is worth noting that, the identified system with GW/FRB association is nevertheless idealistic, but still possible for the next generation of GW detectors, and there could be at least a few hundreds such events detectable per year \cite{Wei:2018cgd}.

In the present work, we will use this upgraded siren to test the constraint ability on  the cosmic anisotropy, assuming that GWs and FRBs have the same progenitor system like BNS or BHNS. The paper is organized as follows. In Sec. \ref{sec:standardsiren}, the upgraded siren with GW/FRB association is reviewed and compared to the usual standard siren when the dipole modulation is concerned. In Sec. \ref{sec:error}, we present the error estimation of the jointed measurements of luminosity distance and dispersion measure. In Sec. \ref{sec:simulations}, the GW/FRB standard siren events are simulated, and the presumed dipole modulation is recovered by adopting the Markov Chain Monte Carlo (MCMC) approach. In Sec. \ref{sec:conclusion}, the results are summarized in conclusions. Throughout this paper, a flat universe is assumed for simplicity, and the geometric unit $c=G=1$ is adopted.

\section{GW/FRB standard siren}\label{sec:standardsiren}
In this section, we will briefly summarize some key points of the GW/FRB standard siren, more details can be found in \cite{Cai:2017aea} and \cite{Wei:2018cgd}. A sensitivity comparison to GW standard siren is given at the end of this section. 

For a GW standard siren event at redshift $z$, the isotropic luminosity distance in a spatially flat FLRW universe
\begin{align}\label{eq:dL}
d_{\rm L}(z)=\frac{1+z}{H_{0}}\int_{0}^{z}\frac{dz^{\prime}}{E(z^{\prime})},
\end{align}
can be directly inferred from the amplitude
\begin{align}\label{eq:Amplitude}
\mathcal{A}=&\frac{1}{d_{\rm L}}\sqrt{F_{+}^{2}(1+\cos^{2}(\iota))^{2}+4F_{\times}^{2}\cos^{2}(\iota)}\times \sqrt{5\pi/96}\pi^{-7/6}\mathcal{M}_{c}^{5/6},
\end{align}
of Fourier transform of GW signal
\begin{align}\label{eq:FTWaveform}
\mathcal{H}(f)=\mathcal{A}f^{-7/6}e^{i\Psi(f)},
\end{align}
whose detector response in the transverse-traceless gauge is
\begin{align}
h(t)=F_{+}(\theta,\phi,\psi)h_{+}(t)+F_{\times}(\theta,\phi,\psi)h_{\times}(t).
\end{align}
Here $E(z)\equiv H(z)/H_{0}=\sqrt{(1-\Omega_{m})+\Omega_{m}(1+z)^{3}}$ is dimensionless Hubble parameter, $H_{0}=100h\,\rm km\,s^{-1}\,Mpc^{-1}$ is the Hubble constant, $\Omega_{m}$ is the matter density parameter today, $\iota$ is the inclination angle between the orbit of the GW source and the line-of-sight of the observer, $\mathcal{M}_{c}$ is the observed chirp mass $\mathcal{M}_{c}=(1+z)M\eta^{3/5}$ with total mass $M=m_{1}+m_{2}$ and symmetric mass ratio $\eta=m_{1}m_{2}/M^{2}$. The phase in Eq. \eqref{eq:FTWaveform} is computed in the post-Newtonian formalism up to 3.5 PN and the specific expression can be found in \cite{Nishizawa:2010xx}. $F_{+,\times}$ are the antenna pattern functions for the two polarizations $h_{+,\times}$, ($\theta,\phi$) are the directional angles of the source in the detector frame, and $\psi$ is the polarization angle. Note that, one can take $\iota=0^{\circ}$ as argued in \cite{Cai:2017aea} and references therein.

On the other hand, the dispersion measure is defined as the observed column density of the free electrons along the line-of-sight \cite{Ioka:2003fr,Inoue:2003ga,Deng:2013aga,Yang:2016zbm},
\begin{align}
\mathrm{DM}=\int\frac{n_{e,z}\mathrm{d}l}{1+z},
\end{align}
which can be measured from the observed time delay between two frequencies of emitted electromagnetic signal \cite{Ioka:2003fr,Inoue:2003ga,Deng:2013aga,Yang:2016zbm},
\begin{align}
\Delta t=\frac{e^2}{2\pi m_e c}\left(\frac{1}{\nu_1^2}-\frac{1}{\nu_2^2}\right)\mathrm{DM}.
\end{align}
When applied to FRB, the observed DM consists of three parts:
\begin{align}
\mathrm{DM}_{\rm obs}=\mathrm{DM}_{\rm MW}+\mathrm{DM}_{\rm IGM}+\frac{\mathrm{DM}_{\rm HG}}{1+z},
\end{align}
which come from the Milky Way (MW), intergalactic medium (IGM), and  the FRB host galaxy (HG), respectively. Once $\mathrm{DM}_{\rm obs}$, $\mathrm{DM}_{\rm MW}$ \cite{Taylor:1993my,Manchester:2004bp,Cordes:2003ik} and $\mathrm{DM}_{\rm HG}$ are determined from observations, one can obtain the observed $\mathrm{DM}_{\rm IGM}$, which contributes most among all with most relevance to cosmological study. The mean value of $\mathrm{DM}_{\rm IGM}$ can be written as \cite{Ioka:2003fr,Inoue:2003ga,Deng:2013aga,Yang:2016zbm}
\begin{align}\label{eq:DM}
\langle \mathrm{DM}_{\rm IGM}(z)\rangle=\frac{3H_0\Omega_bf_{\rm IGM}}{8\pi m_p}\int_{0}^{z}\frac{\chi(z^{\prime})(1+z^{\prime})}{E(z^{\prime})}dz^{\prime},
\end{align}
where the fraction of baryon mass in the IGM $f_{\rm IGM}\simeq0.83$, $\Omega_b$ is the current baryon mass fraction of the universe , and $m_p$ is the mass of proton. 
$\chi(z)=\frac{3}{4}y_1\chi_{\rm e,H}(z)+\frac{1}{8}y_2\chi_{\rm e,He}(z)$, with $y_1\sim1, y_2\sim1$. $\chi_{\rm e,H}(z)$ and $\chi_{\rm e,He}(z)$ are the ionization fractions for H and He, respectively. Since H is essentially fully ionized at $z < 6$, we could take $\chi_{\rm e,H}(z)=1$. For He, however,  it is fully ionized at $z < 3$, so we have the approximate expression for $\chi_{\rm e,He}(z)$, which is given by \cite{Gao:2014iva}
\begin{align}
\chi_{\rm e,He}(z)=\left\{
\begin{array}{lcl}
1,	        &	& {z<3};\\
0.025z^3-0.244z^2+0.513z+1.006, & & {z>3}.
\end{array}\right.
\end{align}
In the rest part of the paper, we will use the notation $\mathrm{DM}_{\rm IGM}$ instead of $\langle \mathrm{DM}_{\rm IGM}\rangle$. 
It was observed in \cite{Wei:2018cgd} that the constraint ability of $d_\mathrm{L}$ or  $\mathrm{DM}_{\rm IGM}$ method alone on the cosmological parameters is limited due to the fact that they both depend on the Hubble constant $H_0$, while $d_\mathrm{L}\cdot \mathrm{DM}_{\rm IGM}$ is independent of $H_0$, as can be seen from Eqs. \eqref{eq:dL} and \eqref{eq:DM}. Note the fact that the local  measurement of $H_0$ is different from that given by the Planck with 4.4 $\sigma$ confidence, which is the so-called ``Hubble Tension'' \cite{Riess:2016jrr,Bernal:2016gxb,Riess:2019cxk}. Thus the inclusion of $\mathrm{DM}_{\rm IGM}$ to $d_\mathrm{L}$ can avoid this problem and exhibit better performance.

One of ways to characterize the cosmic anisotropy is to introduce a dipole modulation with amplitude $g$ and direction $\hat{n}$ on the isotropic luminosity distance,
\begin{align}\label{eq:dLfid}
d_{\rm L}^{\rm fid}(\hat{z})=d_{\rm L}^{\rm iso}(z)[1+g(\hat{n}\cdot\hat{z})],
\end{align}
where the isotropic luminosity distance $d_\mathrm{L}^{\rm iso}(z)$ is calculated according to Eq. \eqref{eq:dL}. Since the definition of dispersion measure also contains the distance element $\mathrm{d}l$ along the line-of-sight, it can be also modified in a similar way as \cite{Qiang:2019zrs} 
\begin{align}\label{eq:DMfid}
\mathrm{DM}_{\rm IGM}^{\rm fid}(\hat{z})= \mathrm{DM}_{\rm IGM}^{\rm iso}(z)[1+g(\hat{n}\cdot\hat{z})].
\end{align}
where the isotropic dispersion measure $\mathrm{DM}_\mathrm{IGM}^{\rm iso}(z)$ is computed according to \eqref{eq:DM}. The dipole direction is given by
\begin{align}\label{eq:n}
\hat{n}=(\cos\phi\sin\theta,\ \sin\phi\sin\theta,\ \cos\theta),
\end{align}
where $\theta\in[0,\pi)$ and $\phi\in[0,2\pi)$. Therefore, the fiducial combination takes the form of
\begin{align}\label{eq:dLDMfid}
d_{\rm L}^{\rm fid}(\hat{z})\cdot \mathrm{DM}_{\rm IGM}^{\rm fid}(\hat{z})=d_{\rm L}^{\rm iso}(z)\cdot \mathrm{DM}_{\rm IGM}^{\rm iso}(z)[1+g(\hat{n}\cdot\hat{z})]^2.
\end{align}

To see that the upgraded siren from GW/FRB performs better than the standard siren from GW alone to test the cosmic anisotropy, one can calculate their relative sensitivity, namely the relative change of measured quantity with respect to the change of the dipole amplitude, 
\begin{align}\label{eq:doublesensitivity}
&\frac{1}{d_{\rm L}^{\rm fid}(\hat{z})\cdot \mathrm{DM}_{\rm IGM}^{\rm fid}(\hat{z})}\frac{\mathrm{d}[d_{\rm L}^{\rm fid}(\hat{z})\cdot \mathrm{DM}_{\rm IGM}^{\rm fid}(\hat{z})]}{\mathrm{d}g} \notag\\
=&2\frac{\hat{n}\cdot\hat{z}}{1+g(\hat{n}\cdot\hat{z})}\notag\\
=&2\frac{1}{d_{\rm L}^{\rm fid}(\hat{z})}\frac{\mathrm{d}[d_{\rm L}^{\rm fid}(\hat{z})]}{\mathrm{d}g}.
\end{align}
The relative sensitivity of the upgraded siren from GW/FRB is twice as large as that of the standard siren from GW alone. Thus we can conclude that $d_{\rm L}\cdot \mathrm{DM}_{\rm IGM}$ is more sensitive than $d_{\rm L}$ with respect to the presumed anisotropy.

\section{Error estimation}\label{sec:error}

ET is a proposed third-generation ground-based GW detector, and the frequency it will cover ranges from 1 to $10^4$ Hz. The exact forms of pattern functions for ET are given by \cite{Zhao:2010sz}
\begin{align}\label{eq:ETPattern}
F_{+}^{(1)}(\theta,\phi,\psi)=&\frac{\sqrt{3}}{2}\bigg[\frac12(1+\cos^{2}(\theta))\cos(2\phi)\cos(2\psi)-\cos(\theta)\sin(2\phi)\sin(2\psi)\bigg];\notag\\
F_{\times}^{(1)}(\theta,\phi,\psi)=&\frac{\sqrt{3}}{2}\bigg[\frac12(1+\cos^{2}(\theta))\cos(2\phi)\sin(2\psi)+\cos(\theta)\sin(2\phi)\cos(2\psi)\bigg],
\end{align}
and the rest of the pattern functions are $F_{+,\times}^{(2)}(\theta,\phi,\psi)=F_{+,\times}^{(1)}(\theta,\phi+2\pi/3,\psi)$ and $F_{+,\times}^{(3)}(\theta,\phi,\psi)=F_{+,\times}^{(1)}(\theta,\phi+4\pi/3,\psi)$, respectively, since the three interferometers align with an angle $60^\circ$ with each other. The error estimations are presented below.

Assuming that the error on the luminosity distance is uncorrelated with errors on the remaining GW parameters, the instrumental error can be estimated with Fisher matrix by
\begin{align}\label{eq:instError1}
\sigma_{d_{\rm L}}^{\rm inst}\simeq\sqrt{\bigg\langle\frac{\partial\mathcal{H}}{\partial d_{\rm L}},\frac{\partial\mathcal{H}}{\partial d_{\rm L}}\bigg\rangle^{-1}}.
\end{align}
Since $\mathcal{H}\propto d_{\rm L}^{-1}$, one could approximate $\sigma_{d_{\rm L}}^{\rm inst}\simeq d_{\rm L}/\rho$. Moreover, a factor of 2 is added to the instrumental error 
\begin{align}\label{eq:instError2}
\sigma_{d_{\rm L}}^{\rm inst}\simeq\frac{2d_{\rm L}}{\rho}
\end{align}
for a conservative estimation \cite{Li:2013lza} when the correlation between $d_{\rm L}$ and $\iota$ is taken into account. Another error to be considered is $\sigma_{d_{\rm L}}^{\rm lens}$ due to the effect of weak lensing, and we assume $\sigma_{d_{\rm L}}^{\rm lens}/d_{\rm L}=0.05z$ as  that in~\cite{Zhao:2010sz}. We therefore take the total uncertainty on the luminosity distance as
\begin{align}\label{eq:ETerrordL}
\sigma_{d_{\rm L}}&=\sqrt{(\sigma_{d_{\rm L}}^{\rm inst})^{2}+(\sigma_{d_{\rm L}}^{\rm lens})^{2}}; \notag\\
&=\sqrt{\bigg(\frac{2d_{\rm L}}{\rho}\bigg)^{2}+(0.05zd_{\rm L})^{2}}.
\end{align}
Thus, the corresponding error for the product $d_\mathrm{L}\cdot \mathrm{DM}_{\rm IGM}$ is given by \cite{Wei:2018cgd}
\begin{align}\label{eq:sigmadLDM}
\sigma_{d_\mathrm{L}\cdot \mathrm{DM}_{\rm IGM}}=\sqrt{(\mathrm{DM}_{\rm IGM}\cdot \sigma_{d_{\mathrm{L}}})^2+(d_\mathrm{L}\cdot \sigma_{\mathrm{DM}_{\rm IGM}})^2}.
\end{align}
In the following section, we will use $d_\mathrm{L}\cdot \mathrm{DM}_{\rm IGM}$ and  Eq. \eqref{eq:sigmadLDM} to simulate the corresponding measurement through Gaussian distribution.

\section{Simulation test}\label{sec:simulations}

Next we simulate a set of GW/FRB events. A GW event is usually claimed when the signal-to-noise ratio (SNR) of the detector network reaches above 8. The combined SNR for the network of $N$ ($N=3$ for ET) independent interferometers is given by
\begin{align}\label{eq:SNR}
\rho=\sqrt{\sum_{i=1}^{N}\langle\mathcal{H}^{(i)},\mathcal{H}^{(i)}\rangle},
\end{align}
where the scalar product is defined as
\begin{align}\label{eq:innerproduct}
\langle a,b\rangle\equiv4\int_{f_{\rm min}}^{f_{\rm max}}\frac{\tilde{a}(f)\tilde{b}^{*}(f)+\tilde{a}^{*}(f)\tilde{b}(f)}{2}\frac{df}{S_{h}(f)}
\end{align}
for given $\tilde{a}(f)$ and $\tilde{b}(f)$ as the Fourier transforms of some functions $a(t)$ and $b(t)$, and $S_{h}(f)$ denotes the one-side noise power spectral density (PSD) characterizing the performance of GW detector. The noise PSD for ET is \cite{Cai:2017aea,Zhao:2010sz}.
\begin{align}\label{eq:ETPSD}
S_{h}(f)=&10^{-50}(2.39\times10^{-27}x^{-15.64}+0.349x^{-2.145}+1.76x^{-0.12}+0.409x^{1.1})^{2}\,\rm Hz^{-1},
\end{align}
where $x=f/f_{p}$ with $f_{p}\equiv100\rm\,Hz$. The lower and upper cutoff frequencies  $f_{\rm min}=1\,\rm Hz$ and $f_{\rm max}=2f_{\rm LSO}$, where the orbit frequency at the last stable orbit $f_{\rm LSO}=1/6^{3/2}2\pi\mathcal{M}_{\rm obs}$ with the observed total mass $M_{\rm obs}=(1+z)M$.

We take the standard $\Lambda$CDM model as our isotropic cosmological model with fiducial values
\begin{align}\label{eq:fiducialpara}
h=0.678,\ \Omega_{\rm m}=0.308,\ \Omega_{\rm b}=0.049
\end{align}
from the Planck 2015 data \cite{Ade:2015xua}. The redshift distribution of the observable sources takes the form of \cite{Zhao:2010sz}
\begin{align}\label{eq:redshift}
P(z)\propto \frac{4\pi d_{\rm c}^{2}(z)R(z)}{(1+z)E(z)},
\end{align}
where $d_{\rm c}$ is the comoving distance defined as $d_{\rm c}(z)\equiv\int_{0}^{z}1/E(z^{\prime})dz^{\prime}$, and $R(z)$ describes the NS-NS merger rate, which is given by \cite{Schneider:2000sg}
\begin{align}\label{eq:mergerrate}
R(z)=\left\{
\begin{array}{lcl}
1+2z,	        &	& {z\leq1};\\
\frac34(5-z),	&	& {1<z<5};\\
0,	            &	& {z\geq5}.
\end{array}\right.
\end{align}

In order to recover the presumed anisotropy with MCMC method, we calculate $\chi^2$ as
\begin{align}\label{eq:chisquare}
\chi^{2}=\sum_{i=1}^{N}\bigg[\frac{d_{\rm L}^{i}\cdot \mathrm{DM}_{\rm IGM}^i-d_{\rm L}^{\rm fid}(\hat{z})\cdot \mathrm{DM}^{\rm fid}_{\rm IGM}(\hat{z})}{\sigma_{d_{\rm L}\cdot \mathrm{DM}_{\rm IGM}}^{i}}\bigg]^{2},
\end{align}
where $N$ denotes the number of data sets, $d_{\rm L}^{i}\cdot \mathrm{DM}_{\rm IGM}^i,\ \sigma_{d_{\rm L}\cdot \mathrm{DM}_{\rm IGM}}^{i}$ are the $i$-th combination of luminosity distance and dispersion measure, and the corresponding error of the simulated data, respectively. 

The specific steps to simulate the mock data of standard siren events are similar with those in~\cite{Cai:2017aea,Cai:2016sby}, which are listed as follows.
\begin{enumerate}
	\item Simulate $N$ GW/FRB datasets, according to the redshift distribution in Eq.~\eqref{eq:redshift}, and the angles $\theta$ and $\phi$ are randomly sampled within the intervals $[0,\pi)$ and $[0,2\pi)$. 
	\item Calculate the isotropic luminosity distance $d_\mathrm{L}^{\rm iso}(z)$ according to Eq. \eqref{eq:dL}, isotropic dispersion measure $\mathrm{DM}_\mathrm{IGM}^{\rm iso}(z)$ from \eqref{eq:DM} and anisotropic fiducial combined $d_\mathrm{L}^\mathrm{fid}(\hat{z})\cdot \mathrm{DM}_{\rm IGM}^{\rm fid}(\hat{z})$ according to Eq. \eqref{eq:dLDMfid}.
	\item Randomly sample the mass of neutron star and black hole within $[1,2]$ $M_{\odot}$ and $[3,10]$ $M_{\odot}$, respectively. Although the current black holes detected by LIGO/Virgo are commonly heavier than $10 M_{\odot}$ (e.g., $30 M_{\odot}$ in GW 150914 \cite{Abbott:2016blz}), we still assume a conservative mass distribution for solar-mass black holes given by \cite{Fryer:1999ht}, as done by many other authors \cite{Cai:2016sby, Zhao:2010sz, Li:2013lza}\footnote{In fact we also did a similar analysis  by assuming  the mass distribution of black holes in $[10,100]$ $M_{\odot}$, we found that the results are qualitatively similar and do not change much. }.  Another reason to assume the black holes in region $[3,10]$ $M_{\odot}$
	is that the FRBs originate from the mergers of binary neutron stars (BNS) or black hole-neutron stars (BHNS).  Assuming the ratio between the number of BHNS and that of BNS is roughly 0.03 \cite{Abadie:2010px}. Evaluate the SNR and the error $\sigma_{d_L\cdot \mathrm{DM}_{\rm IGM}}$.
	\item Simulate $d_{\rm L}^{\rm mea}\cdot \mathrm{DM}_{\rm IGM}^{\rm mea}$ with Gaussian distribution $\mathcal{N}(d_{\rm L}^{\rm fid}\cdot \mathrm{DM}_{\rm IGM}^{\rm fid}, \sigma_{d_{\rm L}\cdot \mathrm{DM}_{\rm IGM}})$ and calculate the $\chi^{2}$ in Eq \eqref{eq:chisquare}. Note that here we do not sample the measurement of $d_{\rm L}^{\rm mea}$ (or $\mathrm{DM}_{\rm IGM}^{\rm mea})$ from $d_{\rm L}^{\rm mea}=\mathcal{N}(d_{\rm L}^{\rm fid}, \sigma_{d_{\rm L}})$ (or $\mathrm{DM}_{\rm IGM}^{\rm mea}=\mathcal{N}(\mathrm{DM}_{\rm IGM}^{\rm fid}, \sigma_{\mathrm{DM}_{\rm IGM}})$) separately, as done in \cite{Wei:2018cgd}. 
	\item Apply the MCMC method to calculate the likelihood function of ($g,\theta,\phi$) and find out the constrained dipole modulation $(g^c,\theta^c,\phi^c)$, which will be compared with the presumed fiducial dipole modulation $(g^f,\theta^f,\phi^f)$.
\end{enumerate}

\begin{figure*}
	\includegraphics[width=0.32\textwidth]{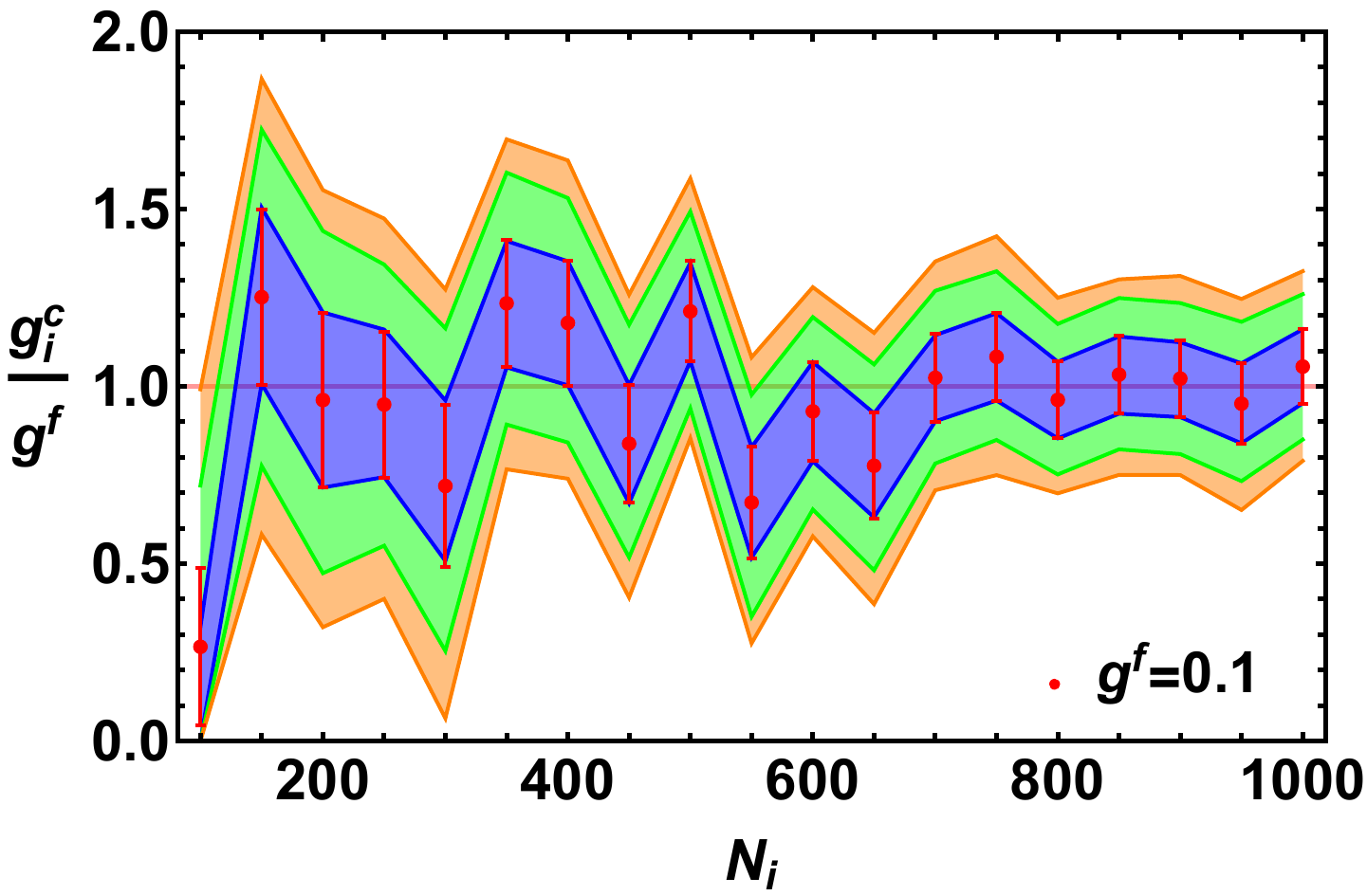}
	\includegraphics[width=0.32\textwidth]{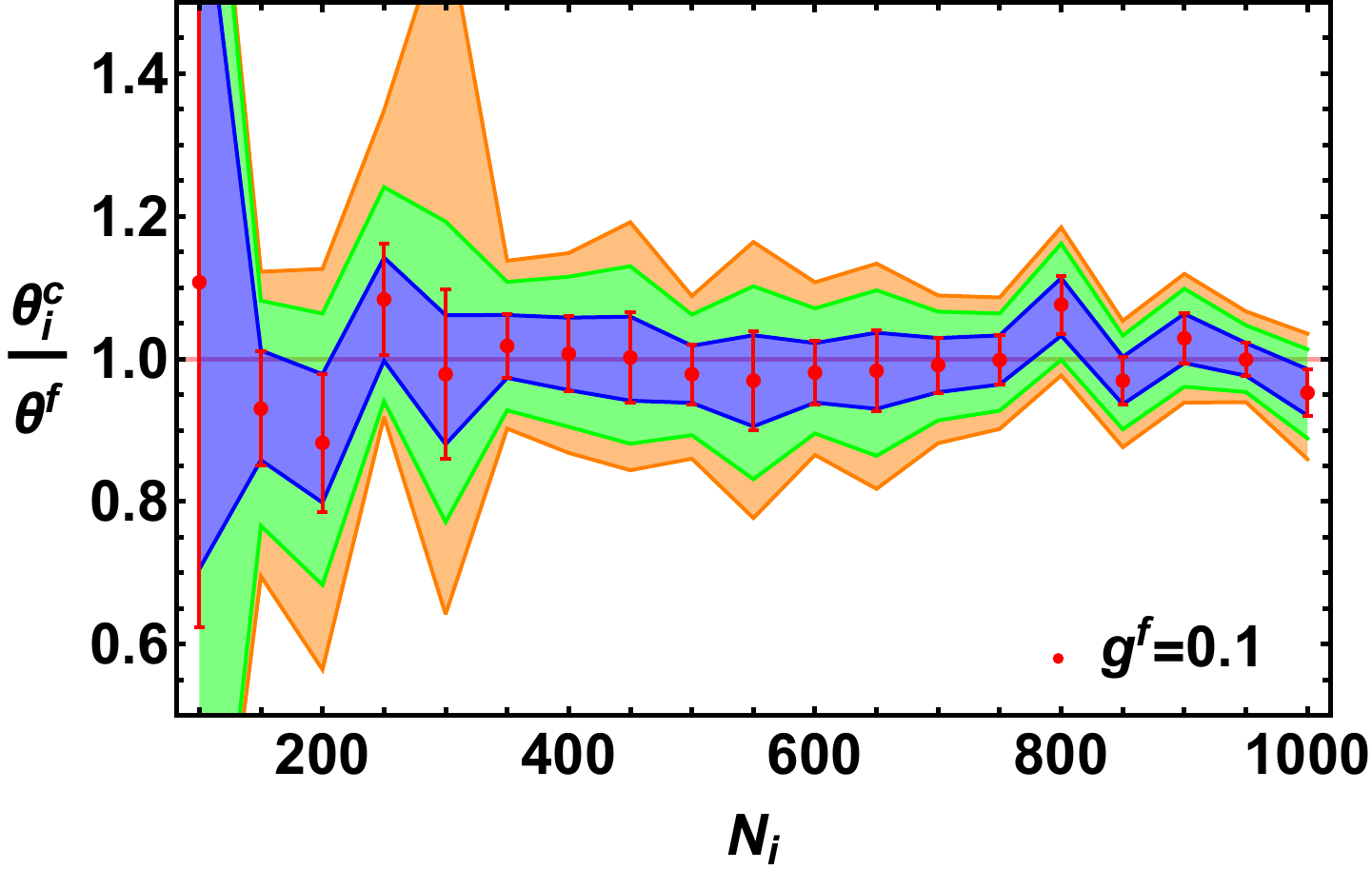}
	\includegraphics[width=0.32\textwidth]{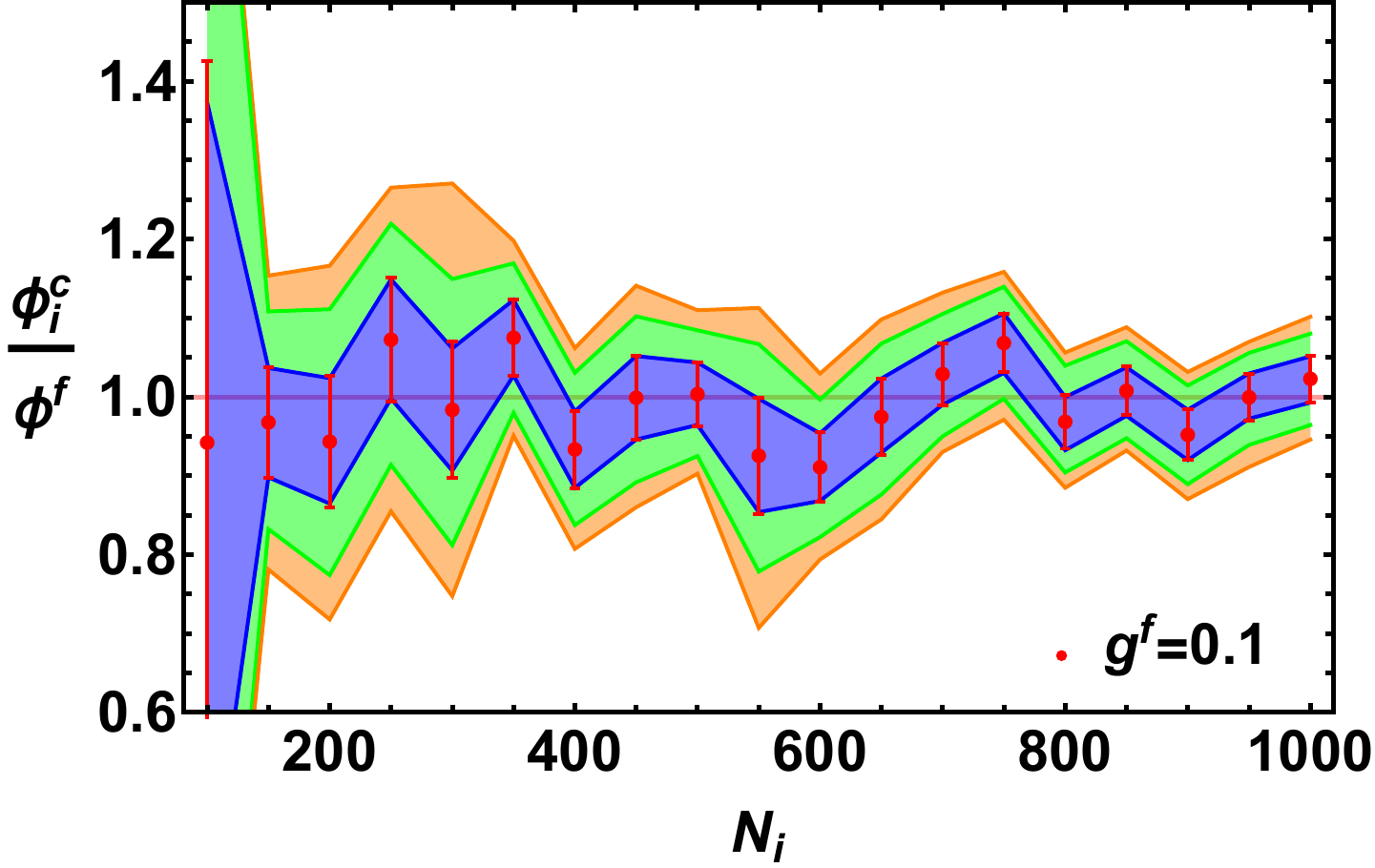}\\
	\includegraphics[width=0.32\textwidth]{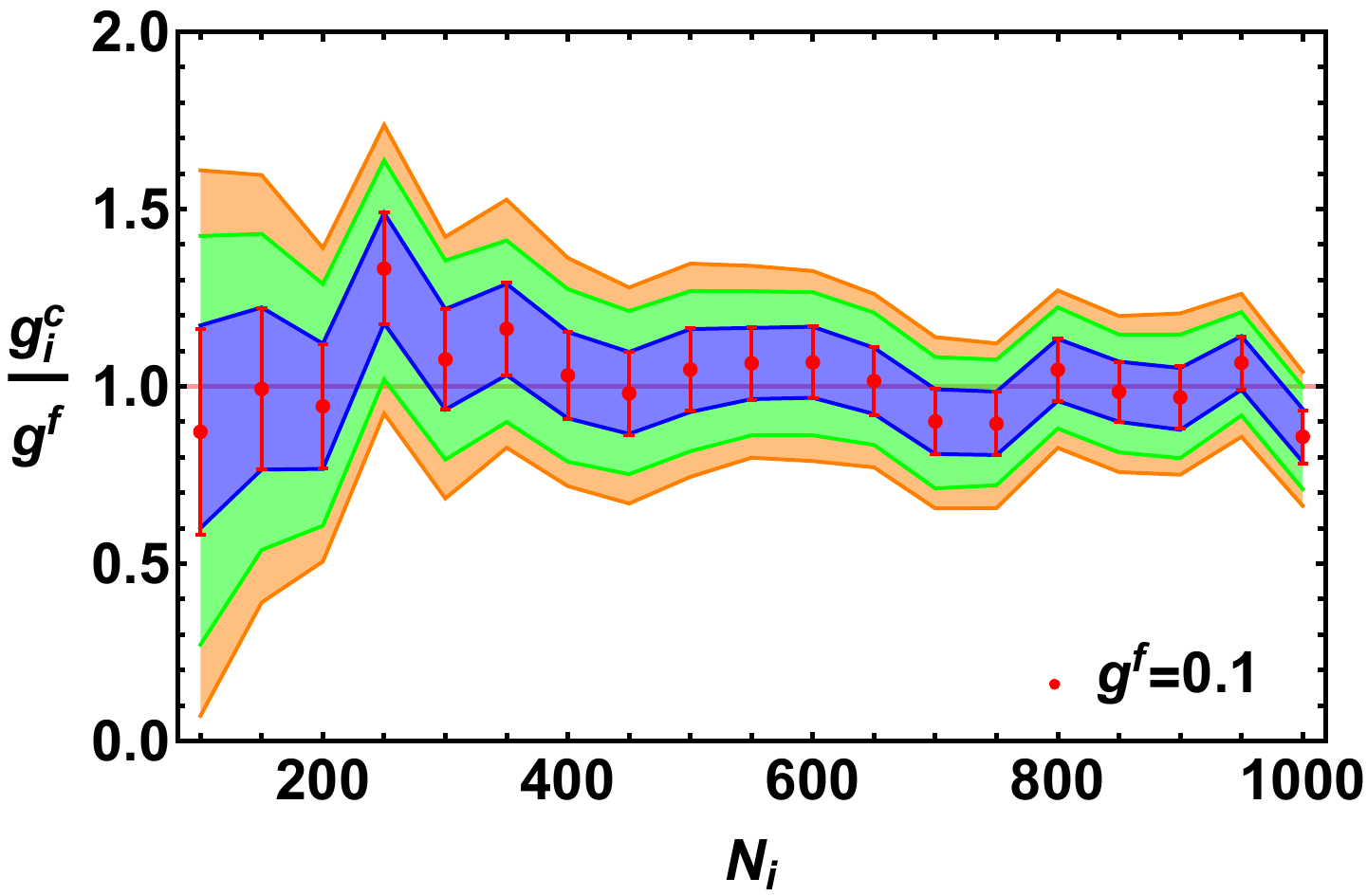}
	\includegraphics[width=0.32\textwidth]{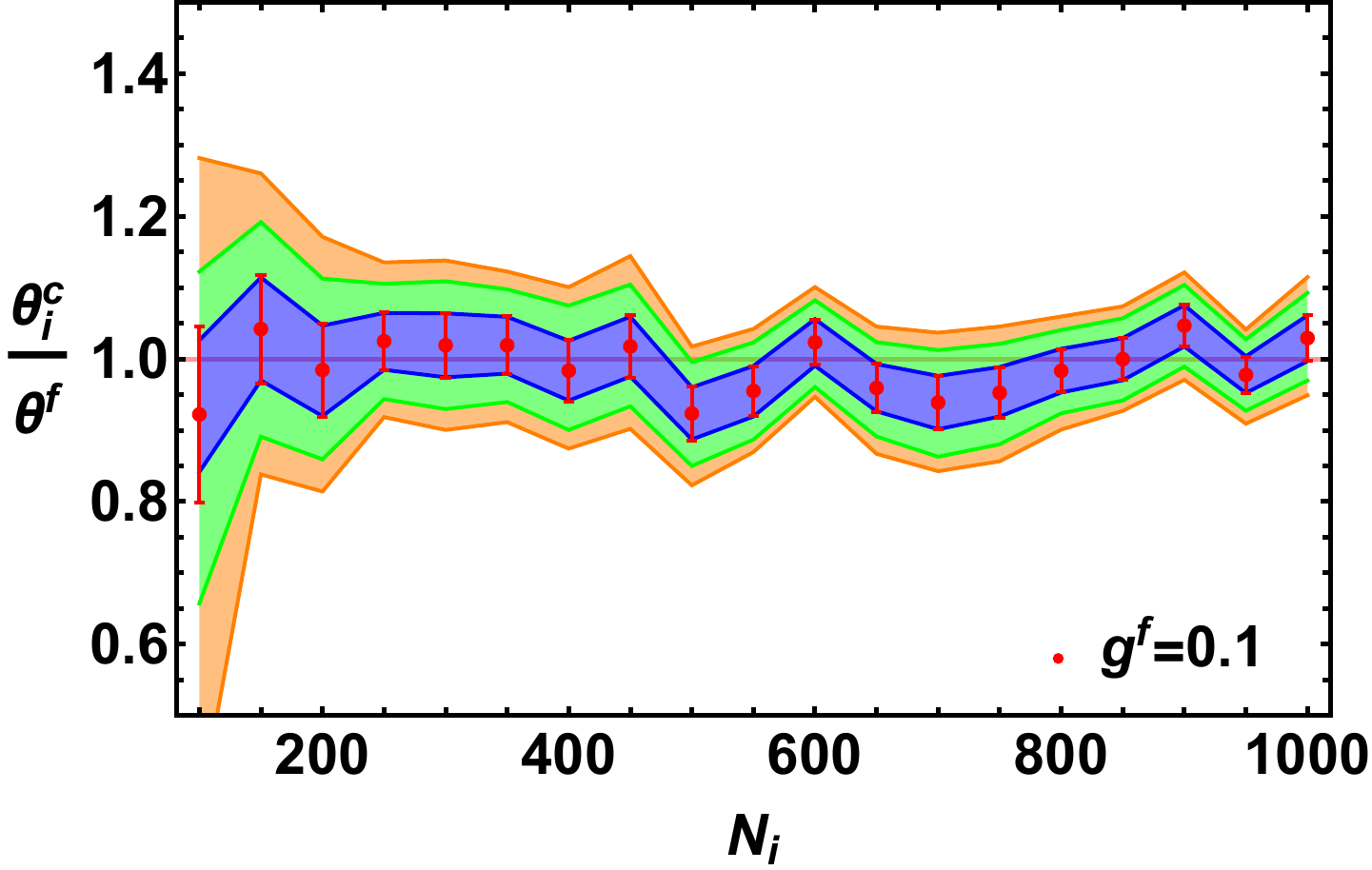}
	\includegraphics[width=0.32\textwidth]{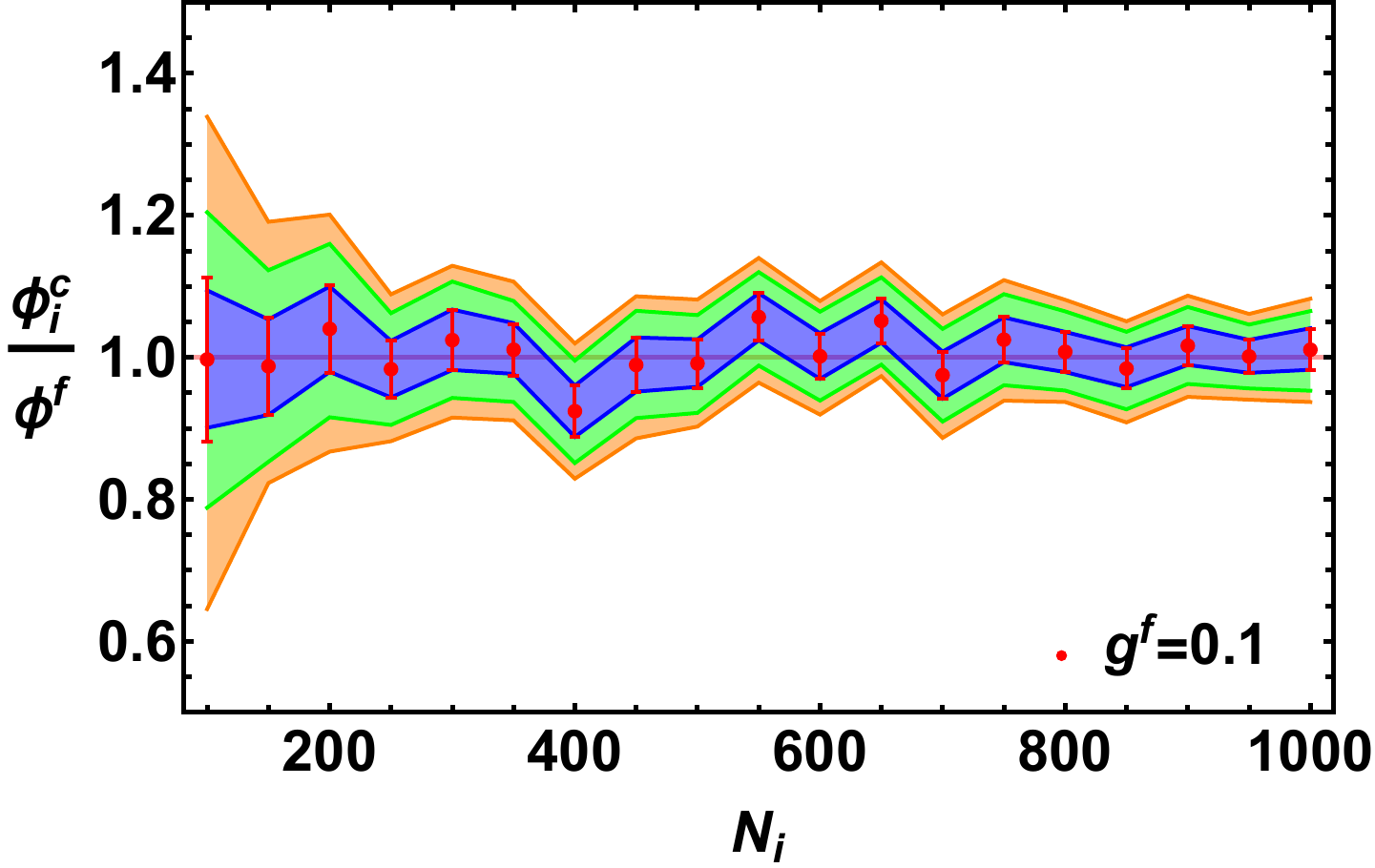}\\
	\includegraphics[width=0.32\textwidth]{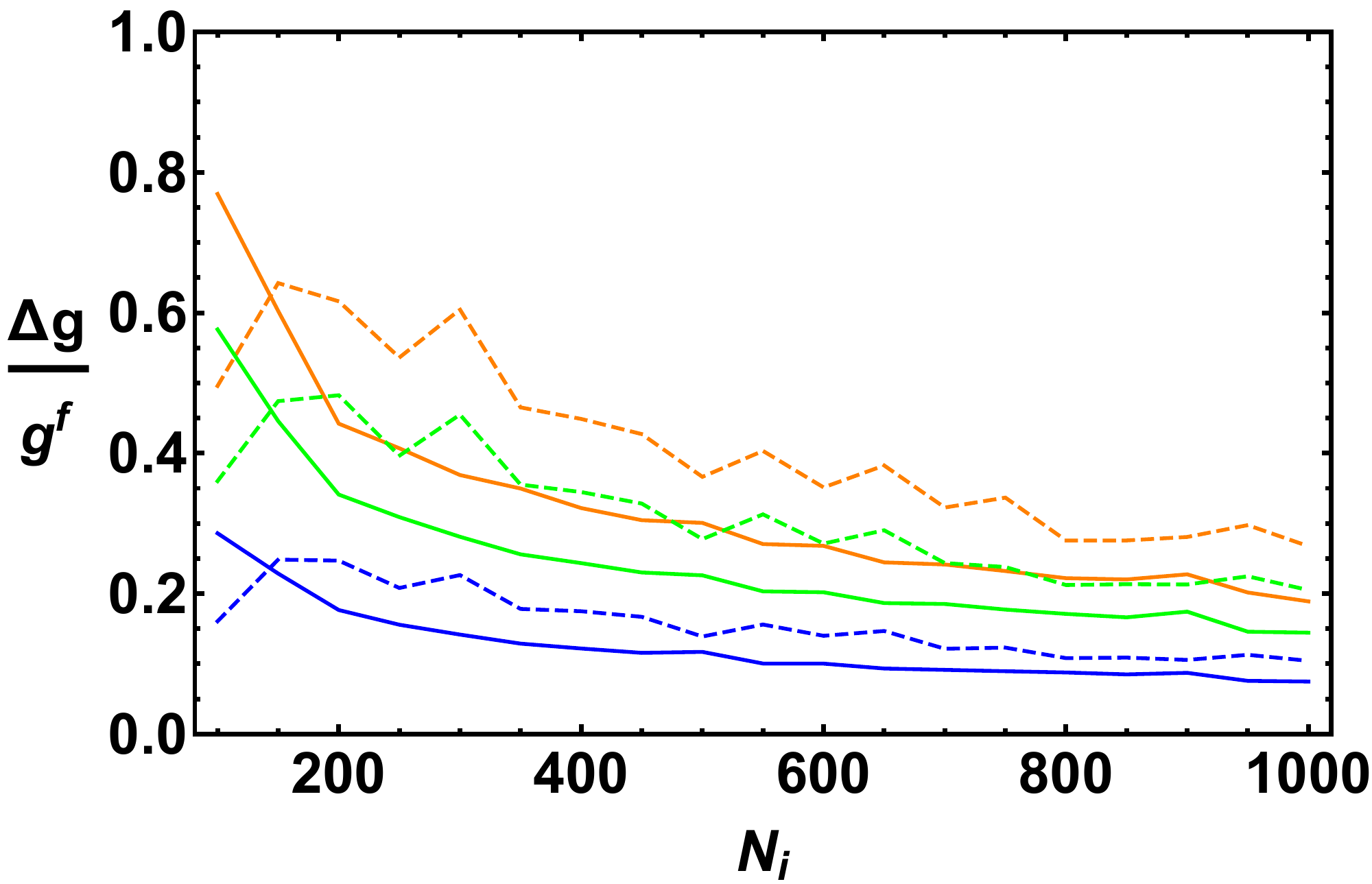}
	\includegraphics[width=0.32\textwidth]{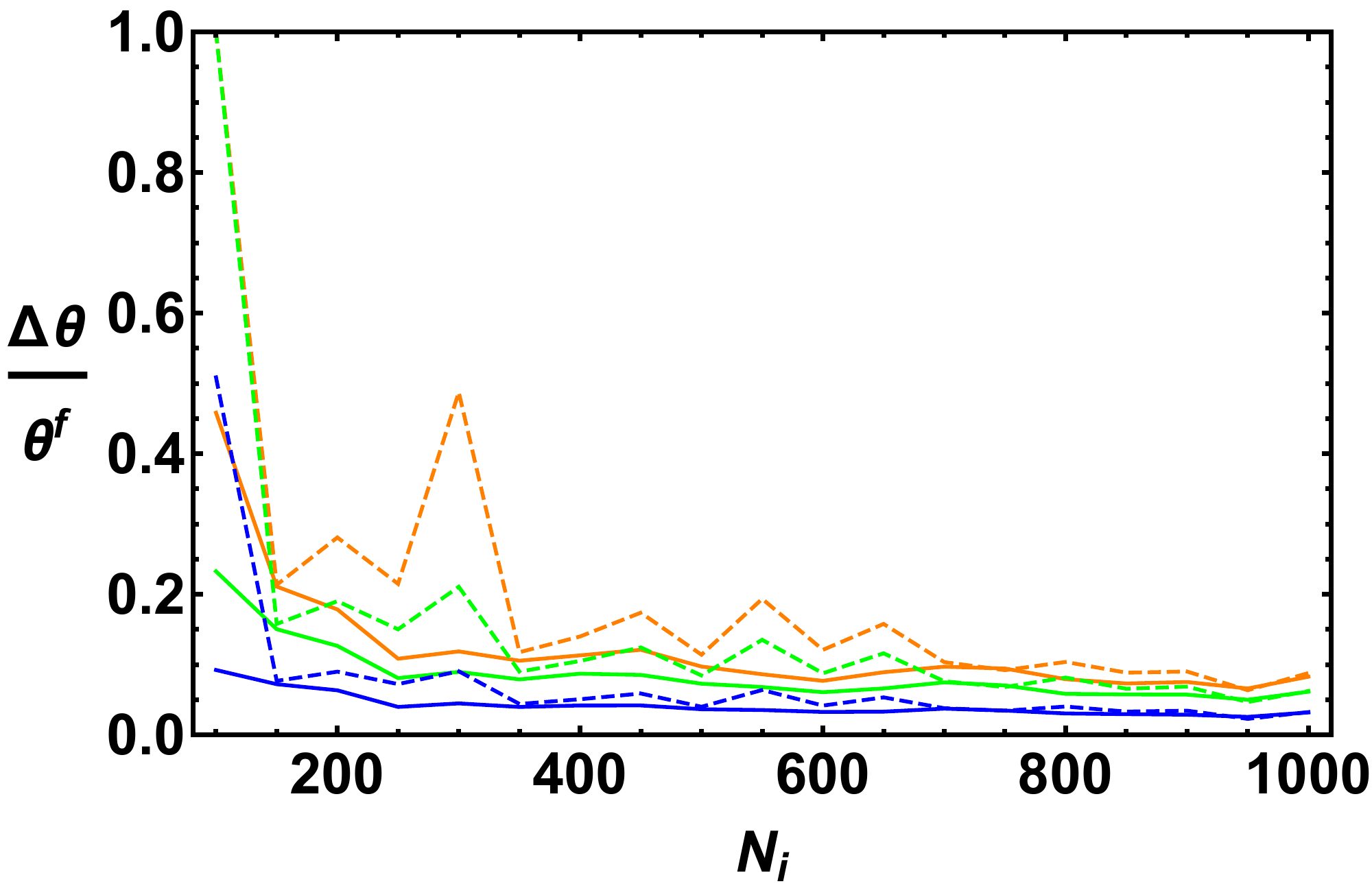}
	\includegraphics[width=0.32\textwidth]{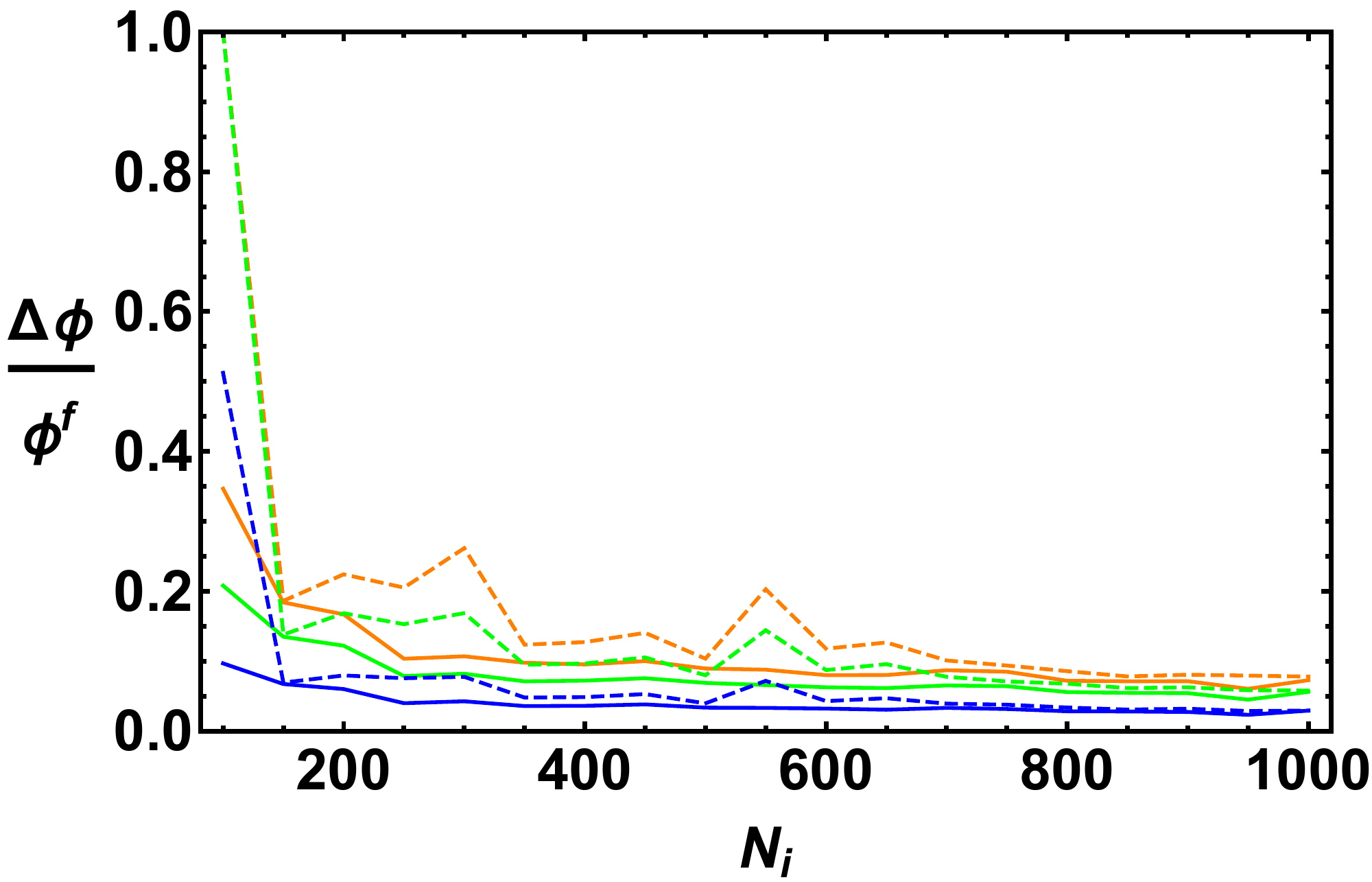}\\
	\caption{The constraint ability of ET with respect to the observed number of standard siren events to recover the given fiducial values of $g^f=0.1$ from GW siren alone (top row) and GW/FRB siren (medium row), whose relative error bar values are compared in the bottom row with dashed lines indicating GW siren and solid lines for GW/FRB siren. The best constrained values divided by the corresponding fiducial values for $g$ (left column), $\theta$ (medium column) and $\phi$ (right column) are labeled by the red dots with standard deviation error bars. The blue/green/orange shaded regions are of 1$\sigma$/2$\sigma$/3$\sigma$ C.L., respectively.}\label{fig:GWandFRBETSigma}
\end{figure*}

\section{Results and discussions}\label{sec:conclusion}

A representative result is illustrated in Fig.~\ref{fig:GWandFRBETSigma} to recover the presumed fiducial value of dipole amplitude $g^f=0.1$ (left column) and the dipole direction angles (medium and right columns). The top-row panels are obtained using GW siren, while the medium-row panels are obtained using GW/FRB siren, of which the error bars are compared in the bottom-row panels with GW siren (dashed) and GW/FRB siren (solid), respectively. As expected theoretically from \eqref{eq:doublesensitivity}, the error bars are all reduced when DM information is included, indicating a better performance of the GW/FRB siren. From Fig.~\ref{fig:GWandFRBETSigma}, we can clearly see that GW sirens are not able to constrain dipole amplitude very well, while the inclusion of FRB could improve the constraint ability significantly. As for dipole directions, GW sirens have already constrained them quite well, so that GW/FRB sirens do not show considerable improvements, especially when increasing $N$ to a larger value.

To summary, we use the combination of GWs and FRBs as the upgraded sirens to investigate the constraint ability on the anisotropy of the universe expansion. This combination takes the advantage of being independent of $H_0$, which reduces the corresponding errors. With the presumed dipole anisotropy, we construct the simulated data of GW events from BNS and BHNS for ET, assuming GWs and FRBs share the same progenitor system. Compared with the result presented in our previous work \cite{Cai:2017aea}, we found a significant improvement on recovering the dipole amplitude. Although there do exist some challenges for this upgraded siren, as discussed in \cite{Wei:2018cgd}, we do believe that it can become a quite promising approach for studying the constraint ability on the cosmic anisotropy when more and more BNS or BHNS events are detected in the future.

\acknowledgments

We want to thank Xue-Wen Liu for helpful discussions and Jun-Jie Wei, He Gao for useful correspondence. This work is supported by the National Natural Science Foundation of China Grants No. 11690022, No. 11435006 and No. 11647601, and by the Strategic Priority Research Program of CAS Grant No. XDB23030100 and by the Key Research Program of Frontier Sciences of CAS. SJW is supported by the postdoctoral scholarship of Tufts University.


\bibliographystyle{JHEP}
\bibliography{ref}

\providecommand{\href}[2]{#2}\begingroup\raggedright\begin{thebibliography}{10}

\bibitem{Schutz:1986gp}
B.~F. Schutz, \emph{{Determining the Hubble Constant from Gravitational Wave
  Observations}}, \href{https://doi.org/10.1038/323310a0}{\emph{Nature}
  {\bfseries 323} (1986) 310}.

\bibitem{Holz:2005df}
D.~E. Holz and S.~A. Hughes, \emph{{Using gravitational-wave standard sirens}},
  \href{https://doi.org/10.1086/431341}{\emph{Astrophys. J.} {\bfseries 629}
  (2005) 15} [\href{https://arxiv.org/abs/astro-ph/0504616}{{\ttfamily
  astro-ph/0504616}}].

\bibitem{TheLIGOScientific:2017qsa}
{\scshape LIGO Scientific, Virgo} collaboration, \emph{{GW170817: Observation
  of Gravitational Waves from a Binary Neutron Star Inspiral}},
  \href{https://doi.org/10.1103/PhysRevLett.119.161101}{\emph{Phys. Rev. Lett.}
  {\bfseries 119} (2017) 161101}
  [\href{https://arxiv.org/abs/1710.05832}{{\ttfamily 1710.05832}}].

\bibitem{Abbott:2017xzu}
{\scshape LIGO Scientific, Virgo, 1M2H, Dark Energy Camera GW-E, DES, DLT40,
  Las Cumbres Observatory, VINROUGE, MASTER} collaboration, \emph{{A
  gravitational-wave standard siren measurement of the Hubble constant}},
  \href{https://doi.org/10.1038/nature24471}{\emph{Nature} {\bfseries 551}
  (2017) 85} [\href{https://arxiv.org/abs/1710.05835}{{\ttfamily 1710.05835}}].

\bibitem{Cai:2017aea}
R.-G. Cai, T.-B. Liu, X.-W. Liu, S.-J. Wang and T.~Yang, \emph{{Probing cosmic
  anisotropy with gravitational waves as standard sirens}},
  \href{https://doi.org/10.1103/PhysRevD.97.103005}{\emph{Phys. Rev.}
  {\bfseries D97} (2018) 103005}
  [\href{https://arxiv.org/abs/1712.00952}{{\ttfamily 1712.00952}}].

\bibitem{Cai:2016sby}
R.-G. Cai and T.~Yang, \emph{{Estimating cosmological parameters by the
  simulated data of gravitational waves from the Einstein Telescope}},
  \href{https://doi.org/10.1103/PhysRevD.95.044024}{\emph{Phys. Rev.}
  {\bfseries D95} (2017) 044024}
  [\href{https://arxiv.org/abs/1608.08008}{{\ttfamily 1608.08008}}].

\bibitem{Cai:2017yww}
R.-G. Cai, N.~Tamanini and T.~Yang, \emph{{Reconstructing the dark sector
  interaction with LISA}},
  \href{https://doi.org/10.1088/1475-7516/2017/05/031}{\emph{JCAP} {\bfseries
  1705} (2017) 031} [\href{https://arxiv.org/abs/1703.07323}{{\ttfamily
  1703.07323}}].

\bibitem{Wang:2018lun}
L.-F. Wang, X.-N. Zhang, J.-F. Zhang and X.~Zhang, \emph{{Impacts of
  gravitational-wave standard siren observation of the Einstein Telescope on
  weighing neutrinos in cosmology}},
  \href{https://doi.org/10.1016/j.physletb.2018.05.027}{\emph{Phys. Lett.}
  {\bfseries B782} (2018) 87}
  [\href{https://arxiv.org/abs/1802.04720}{{\ttfamily 1802.04720}}].

\bibitem{Zhang:2018byx}
X.-N. Zhang, L.-F. Wang, J.-F. Zhang and X.~Zhang, \emph{{Improving
  cosmological parameter estimation with the future gravitational-wave standard
  siren observation from the Einstein Telescope}},
  \href{https://doi.org/10.1103/PhysRevD.99.063510}{\emph{Phys. Rev.}
  {\bfseries D99} (2019) 063510}
  [\href{https://arxiv.org/abs/1804.08379}{{\ttfamily 1804.08379}}].

\bibitem{Wei:2018cov}
J.-J. Wei, \emph{{Model-independent Curvature Determination from
  Gravitational-Wave Standard Sirens and Cosmic Chronometers}},
  \href{https://doi.org/10.3847/1538-4357/aae696}{\emph{Astrophys. J.}
  {\bfseries 868} (2018) 29}
  [\href{https://arxiv.org/abs/1806.09781}{{\ttfamily 1806.09781}}].

\bibitem{Cai:2017cbj}
R.-G. Cai, Z.~Cao, Z.-K. Guo, S.-J. Wang and T.~Yang, \emph{{The
  Gravitational-Wave Physics}},
  \href{https://arxiv.org/abs/1703.00187}{{\ttfamily 1703.00187}}.

\bibitem{Petroff:2016tcr}
E.~Petroff, E.~D. Barr, A.~Jameson, E.~F. Keane, M.~Bailes, M.~Kramer et~al.,
  \emph{{FRBCAT: The Fast Radio Burst Catalogue}},
  \href{https://doi.org/10.1017/pasa.2016.35}{\emph{Publ. Astron. Soc.
  Austral.} {\bfseries 33} (2016) e045}
  [\href{https://arxiv.org/abs/1601.03547}{{\ttfamily 1601.03547}}].

\bibitem{Platts:2018hiy}
E.~Platts, A.~Weltman, A.~Walters, S.~P. Tendulkar, J.~E.~B. Gordin and
  S.~Kandhai, \emph{{A Living Theory Catalogue for Fast Radio Bursts}},
  \href{https://arxiv.org/abs/1810.05836}{{\ttfamily 1810.05836}}.

\bibitem{Lorimer:2007qn}
D.~R. Lorimer, M.~Bailes, M.~A. McLaughlin, D.~J. Narkevic and F.~Crawford,
  \emph{{A bright millisecond radio burst of extragalactic origin}},
  \href{https://doi.org/10.1126/science.1147532}{\emph{Science} {\bfseries 318}
  (2007) 777} [\href{https://arxiv.org/abs/0709.4301}{{\ttfamily 0709.4301}}].

\bibitem{Totani:2013lia}
T.~Totani, \emph{{Cosmological Fast Radio Bursts from Binary Neutron Star
  Mergers}}, \href{https://doi.org/10.1093/pasj/65.5.L12}{\emph{Pub. Astron.
  Soc. Jpn.} {\bfseries 65} (2013) L12}
  [\href{https://arxiv.org/abs/1307.4985}{{\ttfamily 1307.4985}}].

\bibitem{Wang:2016dgs}
J.-S. Wang, Y.-P. Yang, X.-F. Wu, Z.-G. Dai and F.-Y. Wang, \emph{{Fast Radio
  Bursts from the Inspiral of Double Neutron Stars}},
  \href{https://doi.org/10.3847/2041-8205/822/1/L7}{\emph{Astrophys. J.}
  {\bfseries 822} (2016) L7}
  [\href{https://arxiv.org/abs/1603.02014}{{\ttfamily 1603.02014}}].

\bibitem{Yamasaki:2017hdr}
S.~Yamasaki, T.~Totani and K.~Kiuchi, \emph{{Repeating and Non-repeating Fast
  Radio Bursts from Binary Neutron Star Mergers}},
  \href{https://doi.org/10.1093/pasj/psy029}{\emph{Publ. Astron. Soc. Jap.}
  {\bfseries 70} (2018) Publications of the Astronomical Society of Japan,
  Volume 70, Issue 3, 1 June 2018, 39, https://doi.org/10.1093/pasj/psy029}
  [\href{https://arxiv.org/abs/1710.02302}{{\ttfamily 1710.02302}}].

\bibitem{Mingarelli:2015bpo}
C.~M.~F. Mingarelli, J.~Levin and T.~J.~W. Lazio, \emph{{Fast Radio Bursts and
  Radio Transients from Black Hole Batteries}},
  \href{https://doi.org/10.1088/2041-8205/814/2/L20}{\emph{Astrophys. J.}
  {\bfseries 814} (2015) L20}
  [\href{https://arxiv.org/abs/1511.02870}{{\ttfamily 1511.02870}}].

\bibitem{Zhang:2016rli}
B.~Zhang, \emph{{Mergers of Charged Black Holes: Gravitational Wave Events,
  Short Gamma-Ray Bursts, and Fast Radio Bursts}},
  \href{https://doi.org/10.3847/2041-8205/827/2/L31}{\emph{Astrophys. J.}
  {\bfseries 827} (2016) L31}
  [\href{https://arxiv.org/abs/1602.04542}{{\ttfamily 1602.04542}}].

\bibitem{Liu:2016olx}
T.~Liu, G.~E. Romero, M.-L. Liu and A.~Li, \emph{{Fast Radio Bursts and Their
  Gamma-ray or Radio Afterglows as Kerr–newman Black Hole Binaries}},
  \href{https://doi.org/10.3847/0004-637X/826/1/82}{\emph{Astrophys. J.}
  {\bfseries 826} (2016) 82}
  [\href{https://arxiv.org/abs/1602.06907}{{\ttfamily 1602.06907}}].

\bibitem{Falcke:2013xpa}
H.~Falcke and L.~Rezzolla, \emph{{Fast radio bursts: the last sign of
  supramassive neutron stars}},
  \href{https://doi.org/10.1051/0004-6361/201321996}{\emph{Astron. Astrophys.}
  {\bfseries 562} (2014) A137}
  [\href{https://arxiv.org/abs/1307.1409}{{\ttfamily 1307.1409}}].

\bibitem{Nathanail:2017wly}
A.~Nathanail, E.~R. Most and L.~Rezzolla, \emph{{Gravitational collapse to a
  Kerr–Newman black hole}},
  \href{https://doi.org/10.1093/mnrasl/slx035}{\emph{Mon. Not. Roy. Astron.
  Soc.} {\bfseries 469} (2017) L31}
  [\href{https://arxiv.org/abs/1703.03223}{{\ttfamily 1703.03223}}].

\bibitem{Most:2018abt}
E.~R. Most, A.~Nathanail and L.~Rezzolla, \emph{{Electromagnetic emission from
  blitzars and its impact on non-repeating fast radio bursts}},
  \href{https://doi.org/10.3847/1538-4357/aad6ef}{\emph{Astrophys. J.}
  {\bfseries 864} (2018) 117}
  [\href{https://arxiv.org/abs/1801.05705}{{\ttfamily 1801.05705}}].

\bibitem{Qiang:2019zrs}
D.-C. Qiang, H.-K. Deng and H.~Wei, \emph{{Cosmic Anisotropy and Fast Radio
  Bursts}},  \href{https://arxiv.org/abs/1902.03580}{{\ttfamily 1902.03580}}.

\bibitem{Deng:2013aga}
W.~Deng and B.~Zhang, \emph{{Cosmological Implications of Fast Radio
  Burst/Gamma-Ray Burst Associations}},
  \href{https://doi.org/10.1088/2041-8205/783/2/L35}{\emph{Astrophys. J.}
  {\bfseries 783} (2014) L35}
  [\href{https://arxiv.org/abs/1401.0059}{{\ttfamily 1401.0059}}].

\bibitem{Yang:2016zbm}
Y.-P. Yang and B.~Zhang, \emph{{Extracting host galaxy dispersion measure and
  constraining cosmological parameters using fast radio burst data}},
  \href{https://doi.org/10.3847/2041-8205/830/2/L31}{\emph{Astrophys. J.}
  {\bfseries 830} (2016) L31}
  [\href{https://arxiv.org/abs/1608.08154}{{\ttfamily 1608.08154}}].

\bibitem{Gao:2014iva}
H.~Gao, Z.~Li and B.~Zhang, \emph{{Fast Radio Burst/Gamma-Ray Burst
  Cosmography}},
  \href{https://doi.org/10.1088/0004-637X/788/2/189}{\emph{Astrophys. J.}
  {\bfseries 788} (2014) 189}
  [\href{https://arxiv.org/abs/1402.2498}{{\ttfamily 1402.2498}}].

\bibitem{Zhou:2014yta}
B.~Zhou, X.~Li, T.~Wang, Y.-Z. Fan and D.-M. Wei, \emph{{Fast radio bursts as a
  cosmic probe?}},
  \href{https://doi.org/10.1103/PhysRevD.89.107303}{\emph{Phys. Rev.}
  {\bfseries D89} (2014) 107303}
  [\href{https://arxiv.org/abs/1401.2927}{{\ttfamily 1401.2927}}].

\bibitem{Yu:2017beg}
H.~Yu and F.~Y. Wang, \emph{{Measuring the cosmic proper distance from fast
  radio bursts}},
  \href{https://doi.org/10.1051/0004-6361/201731607}{\emph{Astron. Astrophys.}
  {\bfseries 606} (2017) A3}
  [\href{https://arxiv.org/abs/1708.06905}{{\ttfamily 1708.06905}}].

\bibitem{Yang:2017bls}
Y.-P. Yang, R.~Luo, Z.~Li and B.~Zhang, \emph{{Large host-galaxy dispersion
  measure of fast radio bursts}},
  \href{https://doi.org/10.3847/2041-8213/aa6c2e}{\emph{Astrophys. J.}
  {\bfseries 839} (2017) L25}
  [\href{https://arxiv.org/abs/1701.06465}{{\ttfamily 1701.06465}}].

\bibitem{Wei:2018cgd}
J.-J. Wei, X.-F. Wu and H.~Gao, \emph{{Cosmology with Gravitational Wave/Fast
  Radio Burst Associations}},
  \href{https://doi.org/10.3847/2041-8213/aac8e2}{\emph{Astrophys. J.}
  {\bfseries 860} (2018) L7}
  [\href{https://arxiv.org/abs/1805.12265}{{\ttfamily 1805.12265}}].

\bibitem{Li:2017mek}
Z.-X. Li, H.~Gao, X.-H. Ding, G.-J. Wang and B.~Zhang, \emph{{Strongly lensed
  repeating fast radio bursts as precision probes of the universe}},
  \href{https://doi.org/10.1038/s41467-018-06303-0}{\emph{Nature Commun.}
  {\bfseries 9} (2018) 3833}
  [\href{https://arxiv.org/abs/1708.06357}{{\ttfamily 1708.06357}}].

\bibitem{Jaroszynski:2018vgh}
M.~Jaroszynski, \emph{{Fast Radio Bursts and cosmological tests}},
  \href{https://doi.org/10.1093/mnras/sty3529}{\emph{Mon. Not. Roy. Astron.
  Soc.} {\bfseries 484} (2019) 1637}
  [\href{https://arxiv.org/abs/1812.11936}{{\ttfamily 1812.11936}}].

\bibitem{Madhavacheril:2019buy}
M.~S. Madhavacheril, N.~Battaglia, K.~M. Smith and J.~L. Sievers,
  \emph{{Cosmology with kSZ: breaking the optical depth degeneracy with Fast
  Radio Bursts}},  \href{https://arxiv.org/abs/1901.02418}{{\ttfamily
  1901.02418}}.

\bibitem{Wang:2018ydd}
Y.~K. Wang and F.~Y. Wang, \emph{{Lensing of Fast Radio Bursts by Binaries to
  Probe Compact Dark Matter}},
  \href{https://doi.org/10.1051/0004-6361/201731160}{\emph{Astron. Astrophys.}
  {\bfseries 614} (2018) A50}
  [\href{https://arxiv.org/abs/1801.07360}{{\ttfamily 1801.07360}}].

\bibitem{Walters:2017afr}
A.~Walters, A.~Weltman, B.~M. Gaensler, Y.-Z. Ma and A.~Witzemann,
  \emph{{Future Cosmological Constraints from Fast Radio Bursts}},
  \href{https://doi.org/10.3847/1538-4357/aaaf6b}{\emph{Astrophys. J.}
  {\bfseries 856} (2018) 65}
  [\href{https://arxiv.org/abs/1711.11277}{{\ttfamily 1711.11277}}].

\bibitem{Nishizawa:2010xx}
A.~Nishizawa, A.~Taruya and S.~Saito, \emph{{Tracing the redshift evolution of
  Hubble parameter with gravitational-wave standard sirens}},
  \href{https://doi.org/10.1103/PhysRevD.83.084045}{\emph{Phys. Rev.}
  {\bfseries D83} (2011) 084045}
  [\href{https://arxiv.org/abs/1011.5000}{{\ttfamily 1011.5000}}].

\bibitem{Ioka:2003fr}
K.~Ioka, \emph{{Cosmic dispersion measure from gamma-ray burst afterglows:
  probing the reionization history and the burst environment}},
  \href{https://doi.org/10.1086/380598}{\emph{Astrophys. J.} {\bfseries 598}
  (2003) L79} [\href{https://arxiv.org/abs/astro-ph/0309200}{{\ttfamily
  astro-ph/0309200}}].

\bibitem{Inoue:2003ga}
S.~Inoue, \emph{{Probing the cosmic reionization history and local environment
  of gamma-ray bursts through radio dispersion}},
  \href{https://doi.org/10.1111/j.1365-2966.2004.07359.x}{\emph{Mon. Not. Roy.
  Astron. Soc.} {\bfseries 348} (2004) 999}
  [\href{https://arxiv.org/abs/astro-ph/0309364}{{\ttfamily
  astro-ph/0309364}}].

\bibitem{Taylor:1993my}
J.~H. Taylor and J.~M. Cordes, \emph{{Pulsar distances and the galactic
  distribution of free electrons}},
  \href{https://doi.org/10.1086/172870}{\emph{Astrophys. J.} {\bfseries 411}
  (1993) 674}.

\bibitem{Manchester:2004bp}
R.~N. Manchester, G.~B. Hobbs, A.~Teoh and M.~Hobbs, \emph{{The Australia
  Telescope National Facility pulsar catalogue}},
  \href{https://doi.org/10.1086/428488}{\emph{Astron. J.} {\bfseries 129}
  (2005) 1993} [\href{https://arxiv.org/abs/astro-ph/0412641}{{\ttfamily
  astro-ph/0412641}}].

\bibitem{Cordes:2003ik}
J.~M. Cordes and T.~J.~W. Lazio, \emph{{NE2001. 2. Using radio propagation data
  to construct a model for the galactic distribution of free electrons}},
  \href{https://arxiv.org/abs/astro-ph/0301598}{{\ttfamily astro-ph/0301598}}.

\bibitem{Riess:2016jrr}
A.~G. Riess et~al., \emph{{A 2.4\% Determination of the Local Value of the
  Hubble Constant}},
  \href{https://doi.org/10.3847/0004-637X/826/1/56}{\emph{Astrophys. J.}
  {\bfseries 826} (2016) 56}
  [\href{https://arxiv.org/abs/1604.01424}{{\ttfamily 1604.01424}}].

\bibitem{Bernal:2016gxb}
J.~L. Bernal, L.~Verde and A.~G. Riess, \emph{{The trouble with $H_0$}},
  \href{https://doi.org/10.1088/1475-7516/2016/10/019}{\emph{JCAP} {\bfseries
  1610} (2016) 019} [\href{https://arxiv.org/abs/1607.05617}{{\ttfamily
  1607.05617}}].

\bibitem{Riess:2019cxk}
A.~G. Riess, S.~Casertano, W.~Yuan, L.~M. Macri and D.~Scolnic, \emph{{Large
  Magellanic Cloud Cepheid Standards Provide a 1\% Foundation for the
  Determination of the Hubble Constant and Stronger Evidence for Physics beyond
  $\Lambda$CDM}},
  \href{https://doi.org/10.3847/1538-4357/ab1422}{\emph{Astrophys. J.}
  {\bfseries 876} (2019) 85}
  [\href{https://arxiv.org/abs/1903.07603}{{\ttfamily 1903.07603}}].

\bibitem{Zhao:2010sz}
W.~Zhao, C.~Van Den~Broeck, D.~Baskaran and T.~G.~F. Li, \emph{{Determination
  of Dark Energy by the Einstein Telescope: Comparing with CMB, BAO and SNIa
  Observations}}, \href{https://doi.org/10.1103/PhysRevD.83.023005}{\emph{Phys.
  Rev.} {\bfseries D83} (2011) 023005}
  [\href{https://arxiv.org/abs/1009.0206}{{\ttfamily 1009.0206}}].

\bibitem{Li:2013lza}
T.~G.~F. Li, \emph{{Extracting Physics from Gravitational Waves: Testing the
  Strong-field Dynamics of General Relativity and Inferring the Large-scale
  Structure of the Universe}}, Ph.D. thesis, Vrije U., Amsterdam, 2013.

\bibitem{Ade:2015xua}
{\scshape Planck} collaboration, \emph{{Planck 2015 results. XIII. Cosmological
  parameters}},
  \href{https://doi.org/10.1051/0004-6361/201525830}{\emph{Astron. Astrophys.}
  {\bfseries 594} (2016) A13}
  [\href{https://arxiv.org/abs/1502.01589}{{\ttfamily 1502.01589}}].

\bibitem{Schneider:2000sg}
R.~Schneider, V.~Ferrari, S.~Matarrese and S.~F. Portegies~Zwart,
  \emph{{Gravitational waves from cosmological compact binaries}},
  \href{https://doi.org/10.1046/j.1365-8711.2001.04217.x}{\emph{Mon. Not. Roy.
  Astron. Soc.} {\bfseries 324} (2001) 797}
  [\href{https://arxiv.org/abs/astro-ph/0002055}{{\ttfamily
  astro-ph/0002055}}].

\bibitem{Abbott:2016blz}
{\scshape LIGO Scientific, Virgo} collaboration, \emph{{Observation of
  Gravitational Waves from a Binary Black Hole Merger}},
  \href{https://doi.org/10.1103/PhysRevLett.116.061102}{\emph{Phys. Rev. Lett.}
  {\bfseries 116} (2016) 061102}
  [\href{https://arxiv.org/abs/1602.03837}{{\ttfamily 1602.03837}}].

\bibitem{Fryer:1999ht}
C.~L. Fryer and V.~Kalogera, \emph{{Theoretical black hole mass
  distributions}}, \href{https://doi.org/10.1086/321359}{\emph{Astrophys. J.}
  {\bfseries 554} (2001) 548}
  [\href{https://arxiv.org/abs/astro-ph/9911312}{{\ttfamily
  astro-ph/9911312}}].

\bibitem{Abadie:2010px}
{\scshape LIGO Scientific} collaboration, \emph{{Calibration of the LIGO
  Gravitational Wave Detectors in the Fifth Science Run}},
  \href{https://doi.org/10.1016/j.nima.2010.07.089}{\emph{Nucl. Instrum. Meth.}
  {\bfseries A624} (2010) 223}
  [\href{https://arxiv.org/abs/1007.3973}{{\ttfamily 1007.3973}}].

\end{thebibliography}\endgroup
\end{document}